\def\Ob{\Omega_b}

\def\Ol{\Omega_\Lambda}
\def\Om{\Omega_m}

\documentclass[]{aa}
\usepackage{txfonts}

\usepackage{graphicx}

\begin{document}

\title{Constraints on  dark energy evolution}
\titlerunning{}

\author{Luis Ferramacho \inst{1,2} \and
        Alain Blanchard \inst{1} \and
        Yves Zolnierowski \inst{3} \and 
        Alain Riazuelo \inst{4} }

\offprints{Luis Ferramacho}
 
\institute{Laboratoire Astrophysique de Toulouse-Tarbes, CNRS, 14 avenue \'E.Belin, F--31400, Toulouse, France\\ \email{alain.blanchard@ast.obs-mip.fr} 
\and
CENTRA, Departamento de Fisica, Instituto Superior T\'ecnico, Avenida Rovisco Pais, 1 1049-001 Lisboa, Portugal\\ \email{luis.ferramacho@ist.utl.pt}  
 \and LAPP, Universit\'e de Savoie, CNRS/IN2P3 Annecy-le-Vieux, France 9 Chemin de Bellevue, BP 110,
F--74941 Annecy-le-Vieux c\'edex, France\\
\email{zolnierowski@lapp.in2p3.fr} 
\and CNRS,
UMR 7095, Institut d'Astrophysique de Paris, Paris, F-75014 France ;
Universit\'e Pierre et Marie Curie-Paris6, UMR 7095, Paris, F-75014
France \\ \email{riazuelo@iap.fr} }

\date{07/2009}

\abstract{ We investigate a class of dark energy models in which the equation of state undergoes a rapid transition and for which the Hubble  SN~Ia diagram is known to be poorly discriminant. Interestingly enough, we find that  transitions at high redshift can lead to distortion in the correlation function of dark matter at lower redshift.  We therefore use
a combination of the SN~Ia Hubble diagram, Cosmic Microwave Background data and power spectrum from the Sloan Digital Sky Survey Luminous Red Galaxies (SDSS LRG) to constrain the redshift of a possible transition.
We find that the fundamental cosmological parameters are well constrained 
independently of the presence of a transition. Acceptable transitions from an 
equation of state close to $w = 0$ to a value close to $-1$ are strongly rejected at redshifts much higher than 
those for which Large Scale Structure and SN~Ia data are available: the transition redshift can be rejected up to a value as high as 10. We conclude that no preference for a transition appears from present-day data.

}

\maketitle

\section{Introduction}

In  1998 and 1999 two groups collecting far supernovae data  presented evidence for an acceleration of the expansion of the universe, and  consequently for a non-zero cosmological constant,  based on the Hubble diagram of the SN~Ia supernovae (\cite{Riess98}; \cite{Perlmutter}).
Since these times  published supernovae datasets have continuously increased and 
 strengthened the case for an acceleration of the expansion of the universe. 
This has been corroborated by others data sets (\cite{FTH08}; \cite{Blanchard10}).
Although 
the cosmological constant is entirely consistent with the detected acceleration, 
the addition of this term to the theory of general relativity has no known direct motivation. However, the cosmological constant can also be interpreted as arising from the vacuum
 contribution to the energy-momentum tensor. Indeed,  the contributions of the quantum fluctuations of all the fields filling the universe provide a non-zero density to the vacuum and a negative pressure acting exactly as a cosmological constant. However, such anticipated 
 contribution from quantum fluctuations is many orders of magnitude larger than the presently observed value. The low level of the energy scale of the cosmological constant is therefore a mystery. Facing such a difficulty in understanding this discrepancy, one can consider the existence of some unknown mechanism that cancels the contribution from vacuum energy, and look for other physical components in the Universe that can produce an acceleration of the expansion. Motivated by these considerations, \cite{RP88} showed that the presence of a dynamical scalar field, not exactly at rest (otherwise, it would behave exactly 
as a cosmological constant) can actually generate an accelerated expansion. This scenario, known as the quintessence paradigm, has received a lot of attention in recent years after the discovery of the acceleration. Other options have been proposed for its origin.
An attractive idea is the possibility that acceleration appears as a non-linear contribution from inhomogeneities (\cite{Buchert2008}). However, no convincing arguments have been proposed
to show that the actual contribution from these non-linearities can be brought to an observable level much above the naive expectation $<h^2>\sim 10^{-10}$, nor that there is reason why this contribution could have a significant apparent ``negative pressure'' to actually produce an acceleration.
Another more radical option is to modify the equations of the general relativity (\cite{Cognola}).
However,  adding new ``exotic'' components to the universe to obtain the 
acceleration is the most simple solution, and quintessence is natural  in this context: in such  models  dark energy occurs from the late
domination of some scalar field $\phi$. The canonical Lagrangian of such a scalar field is given by 
\begin{equation}
{\cal L}  = \frac{1}{2}(\partial_i \phi)^2 - V(\phi) = X - V(\phi)\;,
\label{eq:Lquin}
\end{equation}
with $X$ being the kinetic energy of the field and $V(\phi)$ the potential. The  equation governing the evolution of this homogeneous field in an expanding universe is
 \begin{equation}
\ddot{\phi}+3H\dot{\phi} = -\frac{dV}{d\phi}(\phi)\;.
\label{eq:vphi}
\end{equation}
The stress-energy tensor has a form identical to that of an ideal fluid with
pressure and density given by the two relations 
\begin{equation}
p = {\cal L} = \frac{1}{2}(\partial_i \phi)^2 - V(\phi)
\label{}
\end{equation}
\begin{equation}
\rho = 2\ X   {\cal L}_{,X} -  {\cal L} = \frac{1}{2}(\partial_i \phi)^2 + V(\phi)\;,
\label{}
\end{equation}
where $_{,X}$ stands for $\displaystyle \frac{\partial} {\partial X}$. For a field which is spatially homogeneous the parameter $w$ is defined from the equation of state
\begin{equation}
w = \frac{p}{\rho} =  \frac{\frac{1}{2}\dot{\phi}^2
-V(\phi)}{\frac{1}{2}\dot{\phi}^2 +V(\phi)}\;,
\label{eq:wx}
\end{equation}
which remains greater than -1, while the sound velocity
\begin{equation}
 c_s^2 = \frac{p_{,X} }{\rho_{, X} } =  \frac{{\cal
L}_{,X}}{{\cal L}_{,X}+2X {\cal L}_{,XX}} = 1
\label{cs}
\end{equation}
is equal to the speed of light. However, more general Lagrangian for the scalar field have been advocated (\cite{Kess}; \cite{Avelino2009}). 
In these models, known as $k$-essence models, the Lagrangian is written in a very general form ${\cal L} = p(X,\phi)$, allowing $w$ to be smaller than -1 and  eventually leading to an arbitrary value of $c_s^2$ 
(\cite{Kess2}). Possible observational consequences of a sound velocity $c_s$
different from the speed of light leads to effects which are below current limits (\cite{cs}). Therefore, for simplicity, we assume below that the sound velocity remains equal to the speed of light and treat $w(z)$ as a phenomenological  function, which should be determined from observations. 

As the damping time scale in Eq. \ref{eq:vphi} is $(3H)^{-1}$, the kinetical energy of the field is 
not expected to change over a time scale much shorter than the Hubble time. 
Therefore, for smooth potentials, $w(z)$ is  expected to vary over a time scale
typical of the Hubble time. For these slowly varying models, the CPL model (\cite{CPL1};\cite{CPL2})
is a
useful approximation that catches the essential feature of many models of varying
dark energy:
\begin{equation}
w_{CPL}(z) = w_0 + w_1 \frac{z}{1+z}\;.
\label{eq:cpl}
\end{equation}
  This model was used to define a figure of merit for a given set of astrophysical observations (\cite{DETF}). 
However, rapid transitions in $w$ at low redshift
 are not well described by such parametrization, and the correct dynamic of dark energy should be taken into account in cosmological parameter estimations (\cite{Vireyetal}). It is 
therefore interesting to investigate a wider class of parametrization. This 
is the purpose of the present work.
In the second section we present the basics of the phenomenology of the quintessence scalar field 
and the parametrization used in this study. In the third section we 
examine  the constraints coming from three commonly used observables, the type Ia supernovae Hubble diagram, the power spectrum  of the cosmological microwave background (CMB) data published by the Wilkinson Microwave Anisotropy Probe (WMAP) collaboration and the power spectrum of Large Red Galaxies (LRG) from the Sloan Digital Sky Survey (SDSS).
 
\section{Dark energy parametrization}
\label{secpara}


\begin{figure*}[t!]
\centering
\includegraphics[width=0.45\textwidth]{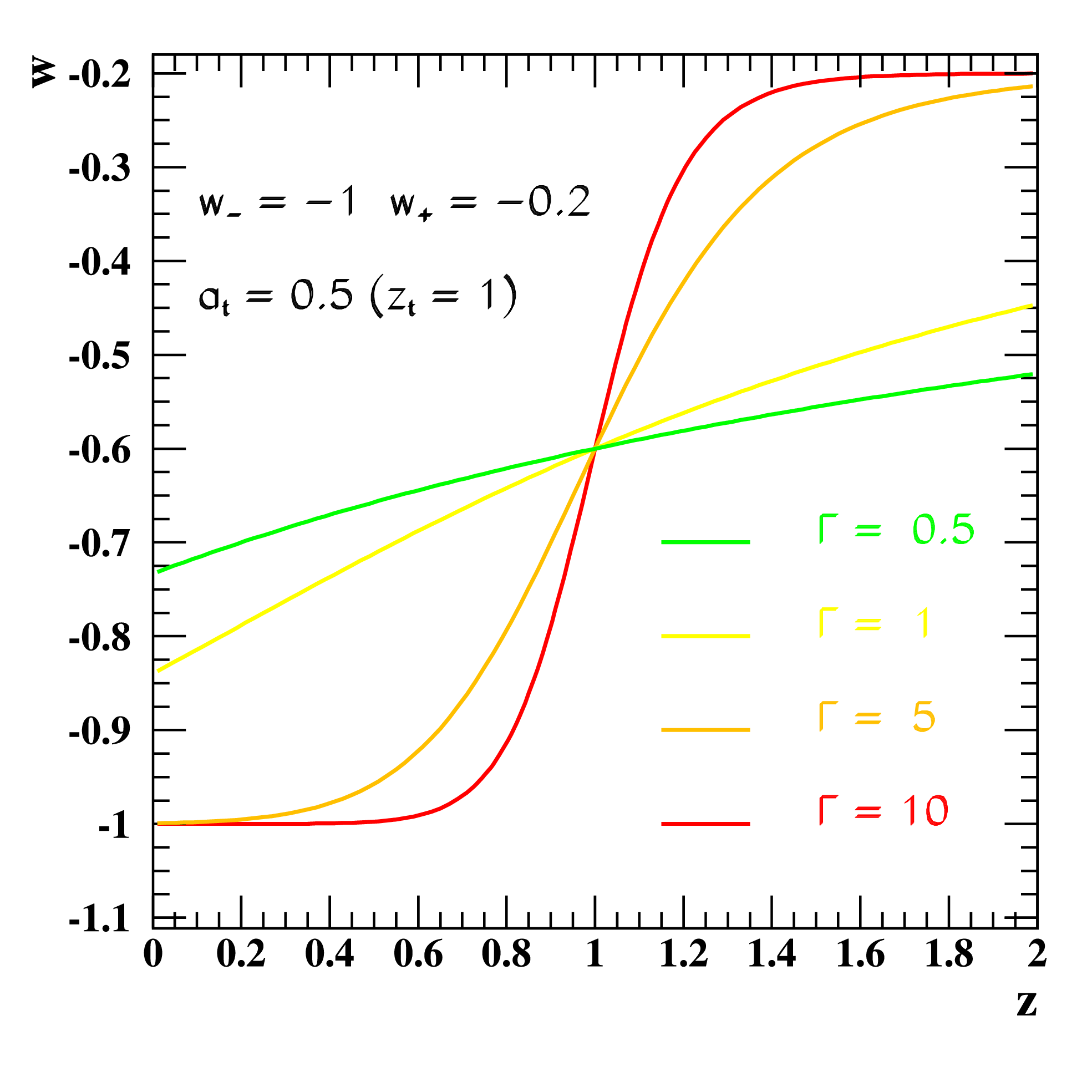}
\includegraphics[width=0.45\textwidth]{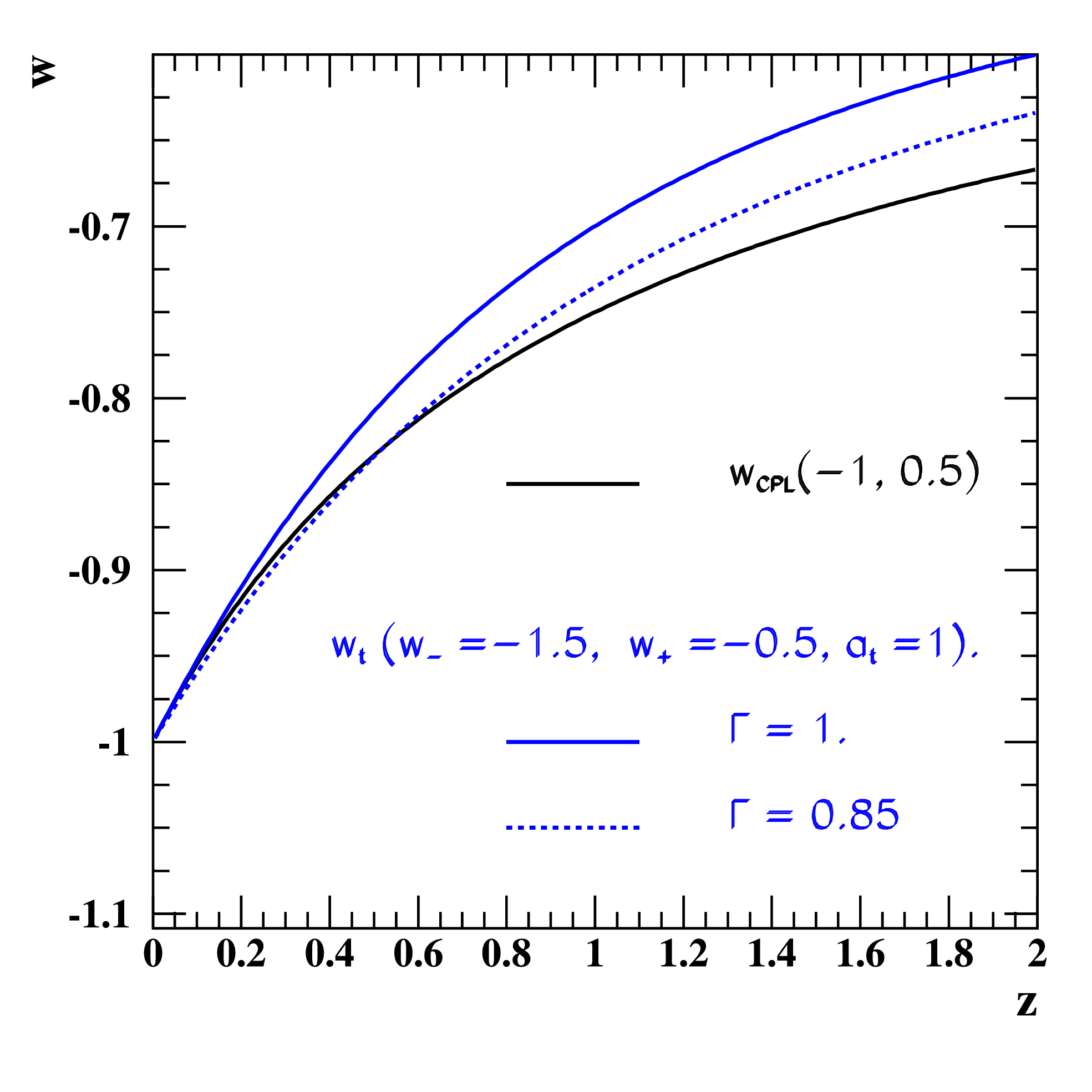}
\caption{ Left: the evolution of the equation of state parameter of dark energy $w(z)$ 
 from Eq. \ref{wqz} is illustrated on this plot for a transition redshift $z_t = 1$ : high values of $\Gamma$ correspond to fast transitions, i.e. whose duration is much shorter than the Hubble time.
Right: comparison between CPL parametrization and Eq. \ref{wqz}.}
\label{fig:trans}
\end{figure*}

For a pressureless  regime like non-relativistic matter one has $w = 0$, for
  radiation $\displaystyle w = \frac{1}{3}$, while for a cosmological constant 
$w = -1$. 
As we mentioned above, rapid transitions in $w$ at non vanishing  redshift
 are not well described by CPL parametrization (\cite{LindenVirey2008}). It is 
therefore interesting to investigate a wider class of parametrization. A convenient
parametrization is (\cite{2005PhRvD..72d3509L}; \cite{Douspis2008})
\begin{equation}
\label{wqz}
w_t(z) = \frac{1}{2} (w_+ + w_-) -
\frac{1}{2} (w_+ - w_-) \tanh \left(\Gamma
\log\left(\frac{1+z_t}{1+z}\right) \right) .
\end{equation}
The equation of state parameter $w_t$ evolves from $w_t(z=+\infty) = w_+$ to $w_t(z=-1)
= w_-$, the transition occurs at redshift $z_t$ or equivalently $a_t=1/(1+z_t)$. The parameter $\Gamma$ figures  
the hardness of the transition: high values of $\Gamma$ correspond to rapid 
transitions, i.e. transitions with a shorter duration than the Hubble time at
 $z_t$, while  $\Gamma \sim 1$ corresponds to transitions occurring over a period
similar to the Hubble time. The above parametric form may hide complexities :
$w < -1$ (\cite{Caldwell2002}) models might be pathological (\cite{wmoins1}). 
We try to keep our investigation as open as possible and do not 
add theoretical constraints on the kinds of models we investigate. For instance, 
modified gravity theories might lead to an evolution in w(z) which crosses the 
phantom limit of $w = -1$ (\cite{wcross1}; \cite{wcross2}). We therefore do not impose any restriction on the classes of models we investigate in terms of the values of $w_+$ and $w_-$.

With an appropriate choice of the parameter $\Gamma $, this
parametrization is similar to the CPL parametrization (Eq. \ref{eq:cpl}) for $z\geq0$:
\begin{equation}
w_0 = \frac{1}{2} (w_+ + w_-) \\
w_1 = \frac{1}{2} (w_+ - w_-)\;. \\
\end{equation}
Requesting that the derivative of $w_t$ function at $z=0$ is the same as the derivative $w_{CPL}$ function ($w_t'(z=0) = w_{CPL}^{'}(z=0)$) imposes 
 $\Gamma = 1$ and $a_t=1$. However, slightly different values of $\Gamma$ might lead to a $w_t(z)$ 
closer to $w(z)$ over a wide  redshift range. For instance,
with  $\Gamma \sim 0.85$, 
$w(z)$ is very close to the CPL parametrization in the redshift range $0$ to $1$ (see Fig. \ref{fig:trans}). The above parametrization  provides therefore a nice generalization, comprising a close approximation of the CPL case, but also allowing the investigation of more rapid transitions.

\section{Constraints from present-day cosmological data}
\label{secanalysis}

Modifications of the equation of state of dark energy obviously lead to 
modification of the expansion rate through the Friedmann-Lema\^ \i tre equation
 \begin{equation}
\ddot{R} = -4\pi G (\rho_m+\rho_X(1+3w(z))\;.
\end{equation}
This has direct consequences on geometrical tests. Another important aspect comes from the fact that  the growth rate of fluctuations in the linear regime is modified because the nature of the background is changed. For $w_+ \sim 0$ the contribution of dark energy can be non-negligible at high redshift and therefore may leave specific imprints in the CMB and in the matter power spectrum. These imprints may not be reducible to a single geometrical factor (\cite {FBZ}).

\subsection{Supernovae Hubble diagram}

In a preceding paper (\cite{Douspis2008}) we presented some evidence that supernovae alone cannot reject a sharp transition between two values of $w$: $w_+$ and $w_-$ above a transition redshift $z_t(z)$ =  0.2 at 2 sigma using a compilation of supernovae recently published  (\cite{Davis2007}). We chose for the two extreme values of $w$, -0.2 for $w_+$ and -1 for  $w_-$. The choice of the second value was linked to the observation that the value of $w$ now seems very close to -1, and the  choice of $w_-$ was discussed in detail in the article. 
There was also some surprising but weak evidence  that supernovae prefer a sharp transition at low redshift (\cite{2002MNRAS.336.1217B}).
The number of published SN~Ia samples has increased in recent years. In 2008, Kowalski and coworkers (\cite{Kowalski2008})
have applied a homogeneous treatment to 13 samples of SN1a published to reduce the possible systematics  due to the analysis done under different conditions and by  different teams as much as possible. 
The tendency of SN~Ia sample to prefer a transition has still be claimed 
to be  present in this last homogeneous supernovae data set  (\cite{2009arXiv0903.5141S}). This provides us with  motivation for more detailed investigations.

\subsection{Constraints from the CMB}

In order to use measurements of the CMB fluctuations to constrain models with a rapid transition in  dark 
energy we need to implement these models in a code which computes the CMB 
fluctuations angular spectrum. We therefore use a modified version of CAMB in which our dark energy parametrization has been implemented (see \cite{Douspis2008} for details on this implementation). As the value of the dark energy density is analytically known in this model, this implementation is straightforward, something 
which might not be the case if it has to be evaluated through numerical integration (\cite{Corasantini2004}), as is generally the case. The sound speed was kept constant. We did not use the reduced quantity $R$, which is closely related to the angular distance to the CMB, and which is the most changed quantity when adding a cosmological constant  (\cite{Blanchard84}; \cite{EfstathiouBond}). We used the 5-year release (\cite{wmap5}) and the ACBAR data (\cite{ACBAR})  for additional stringent constraint on small scales.

\subsection{The effects of a transition model for dark energy in large scale structure data}
\label{secliv}

Before trying to constrain any fast transition model using all data, it is interesting to examine how these models affect the predicted matter distribution at low redshifts. For this purpose, we used the correlation function data of \cite{E05}, EO5 hereafter, which provide a helpful visualization of baryonic acoustic oscillations, since they appear as a well defined feature, the so-called acoustic  peak in this correlation function corresponding to the oscillations in the power spectrum (BAO for baryonic acoustic oscillation) .  

We applied the procedure for determining the expected correlation function as described in Ferramacho et al. (2009), with a modified CAMB routine that comprises our varying dark energy parametrization (\cite{Douspis2008}). As discussed in Ferramacho et al. (2009), non-linear corrections and scale dependent bias play an important role in the final constraints 
obtained from LSS distribution and BAO. We have thus applied the same prescriptions for these corrections, but taking some caution. For instance, we did not include the modeling that accounts for the suppression of the baryonic acoustic peak, since it uses the no-wiggle approximation from \cite{EisensteinHU}, which comprises fitting formulas derived for a standard $\Lambda$CDM cosmology. 
We checked however that correcting for this effect does not change the overall constraints on $w$ and the density parameters, except for $\Omega_b$, which is expected since the amplitude of the acoustic peak depends mostly on the baryonic content of the universe rather than on any other quantity. In order to perform these corrections at low scales with less model dependence, it is also preferable to choose a more general correction model than the E05 approach. For this purpose, a Q-model (\cite{Cole05}) was used with $b$ and $Q$ as free parameters, which seems  to be more appropriate for dark energy models with dynamical equation of state.

\begin{figure}[t!]
\includegraphics[width=0.45\textwidth]{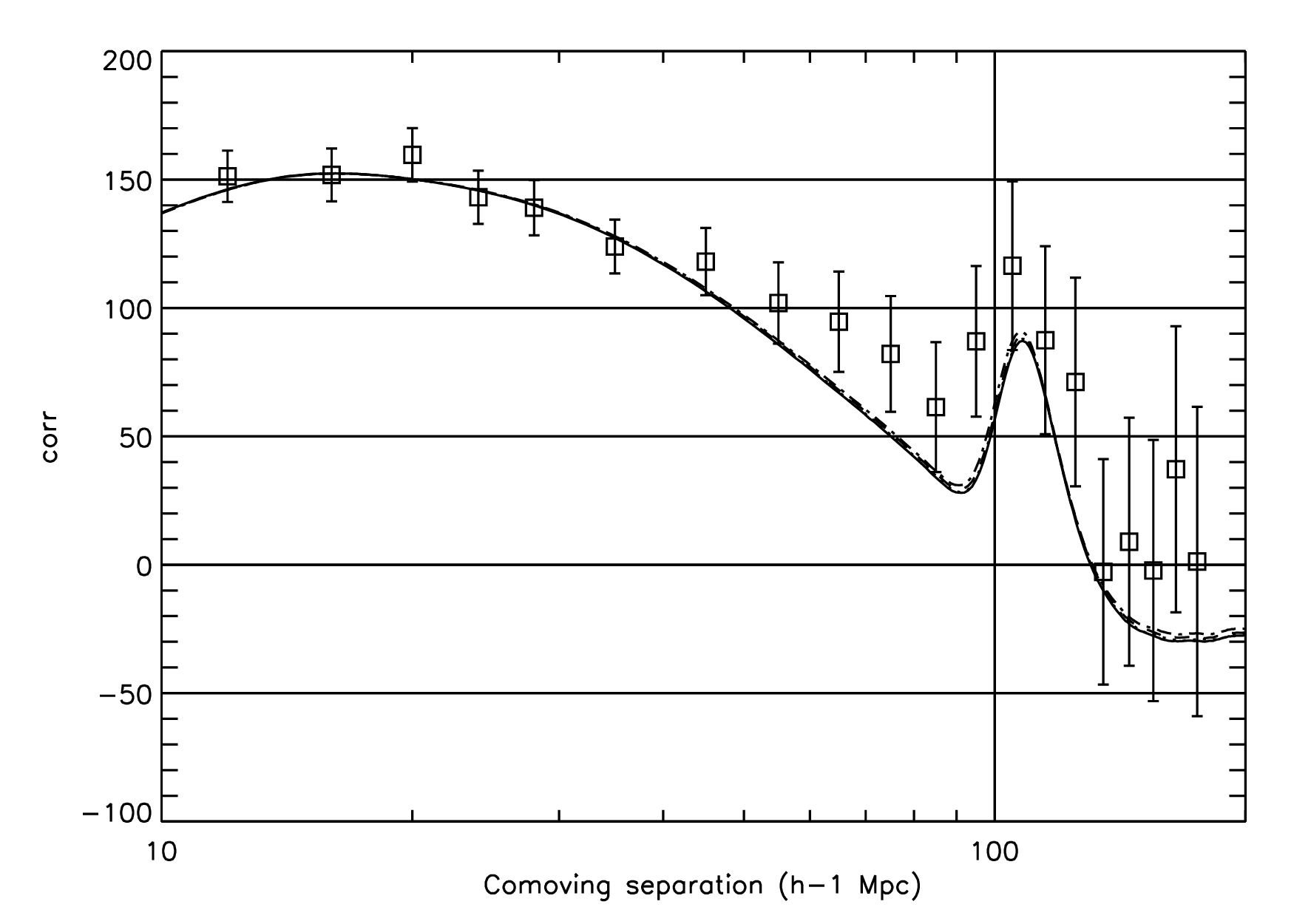}\\
\includegraphics[width=0.45\textwidth]{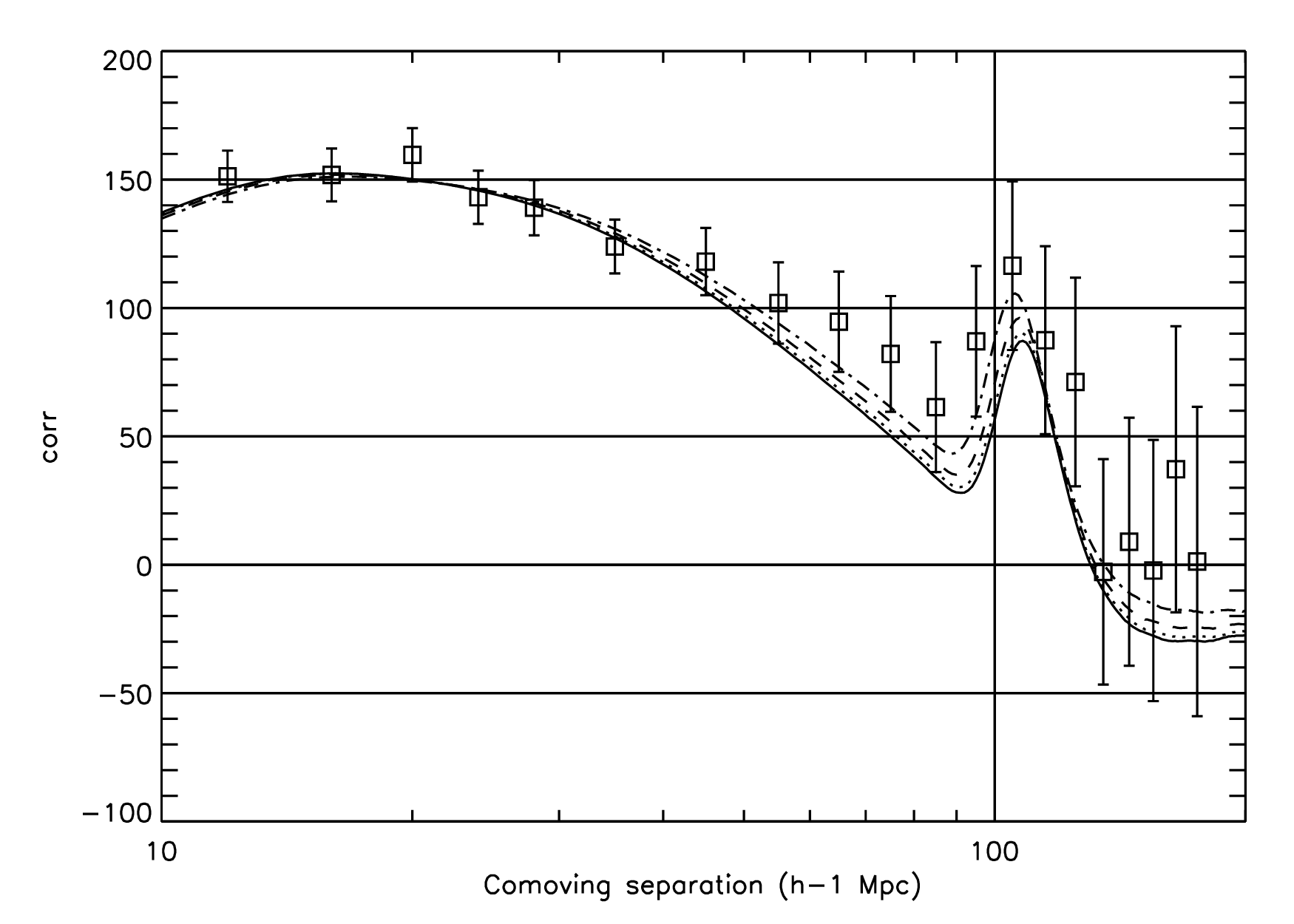}
\includegraphics[width=0.45\textwidth]{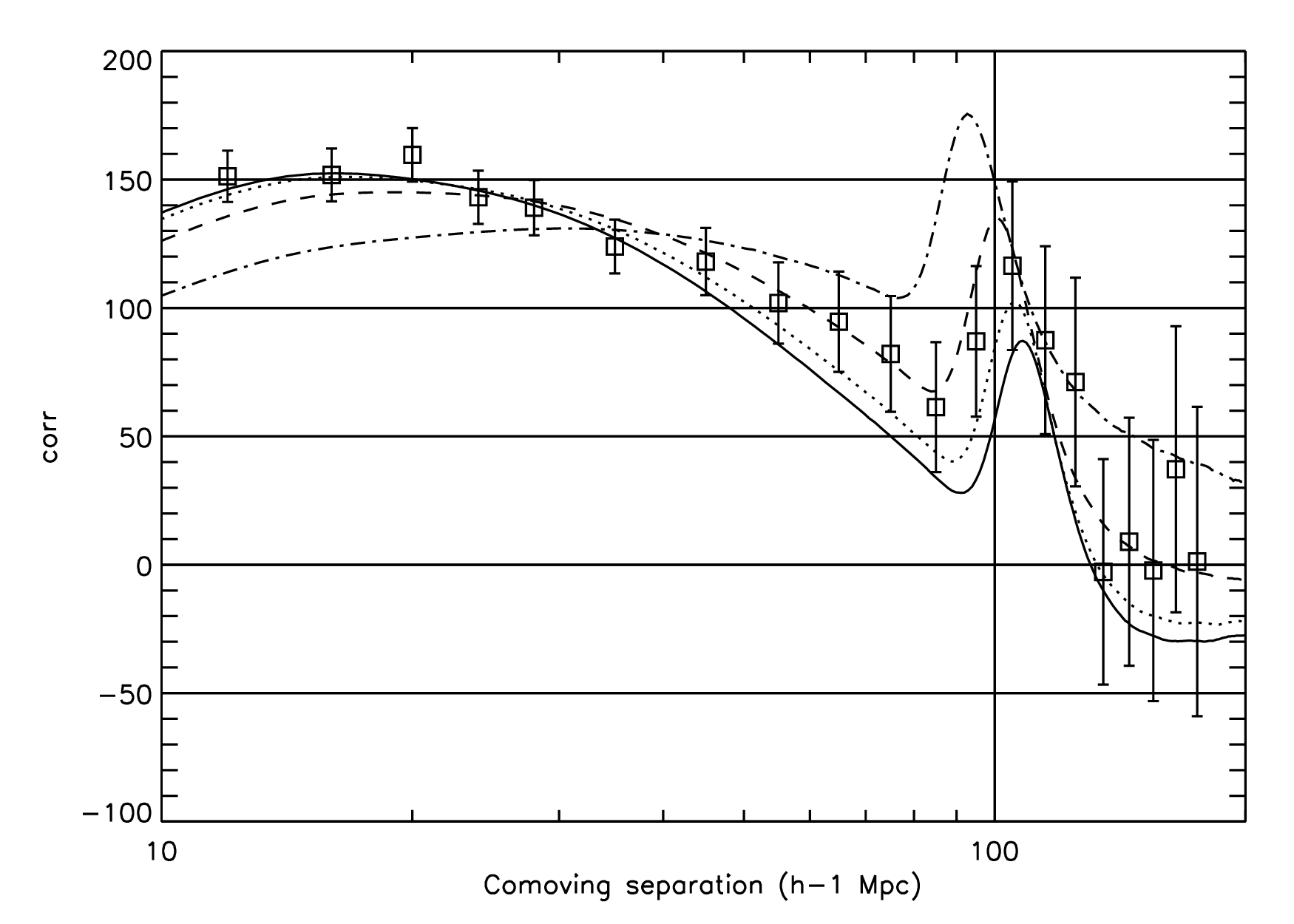}
\caption{Expected shape for the correlation function in units of $\xi r^2$, assuming rapid (step-like) transitions with $\Gamma = 5$ in the dark energy equation of state at different redshifts, with $w_-=-1$. \emph{Upper panel:} $w_+=-0.2$. \emph{Middle panel:} $w_+=-0.1$. \emph{Lower panel:} $w_+=0$. Plotted curves correspond to different values for the transition redshift: $z_t=0.5$ (\emph{dashed-dot}), $z_t=1$ (\emph{dashed}),  $z_t=2$ (\emph{dot}) and  $z_t=+\infty$ (\emph{straight}).}
\label{fig:xi}
\end{figure}

 In Fig. \ref{fig:xi} we illustrate  the effect of fast transitions in the correlation function at low redshift by considering three different values for $w_+$ and setting $w_-=-1$. In each plot, the effect of changing the transition epoch ($a_t$) is shown. The remaining cosmological parameters were set to the best values obtained in Ferramacho et al. (2009) when considering a free but constant value for $w$. The bias parameter was marginalized over in all the plots for a better comparison of the effects in the overall shape. Matter fluctuation amplitude, measured by $\sigma_8$, might therefore be different for the different models.


Transitions between $w_+=-0.2$ and $w_-=-1$ do not change the overall shape of the correlation function, even if we consider transitions at relatively low redshifts. In this case, the predicted matter distribution with the adopted model for dark energy cannot be distinguished from the same distribution obtained with a standard $\Lambda$CDM model. However, when we let $w_+$ be closer to $0$, some interesting effects appear in the predicted correlation function, noticeably at intermediate and large scales. We observe then a relative increase of the power at these scales, which boosts the amplitude of the baryonic peak (to a small degree when $w_+=-0.1$ and much larger when $w_+=0$) and shifts its position, and furthermore the global shape of the correlation function is modified. These effects show that the matter distribution is sensitive to fast and strong variations in the dark energy equation of state and provides in principle a good probe to test such models. Also, in the view of these results changes in the correlation function for these models cannot be fully reproduced with a single geometrical factor, as for instance the $A$ parameter from E05.
So, the full shape of the correlation function (or of the matter power spectrum) must be taken into account when studying models that have an effect on both the correlation function form and the position of the baryonic peak due to other effects than angular diameter distance changes. The precision of the SDSS LRG measurements on the correlation function is not high, but as an example 
 the model with $w_+=-0.1$ and a transition redshift fixed at $z_t=0.5$ has 
a $\chi^2$ reduced of 2 in comparison with 
the concordance model, suggesting that a transition is preferred at 1 $\sigma$ using only the correlation function data.

That transitions in $w$ at redshifts as high as 1 could leave an imprint on the correlation function is an interesting issue requiring a further investigation.
Below, we look into the impact on these models of the joint analysis of data from galaxy distribution, CMB anisotropies and SN~Ia. 

\subsection{Hubble time transitions - CPL approximation}

Let us start our analysis  by taking the case of transitions occurring at a rate typical of the Hubble time. As seen above, with $\Gamma \sim 1$, the parametrization provided by Eq. \ref{wqz} becomes almost equivalent to the CPL parametrization.
It is interesting to compare the effect of considering $\Gamma=1$, which fulfills the condition $w_t^{'}(z=0)=w_1$, as requested when deriving the CPL parametrization, and a slightly lower value for $\Gamma$ like 0.85, which provides a better agreement for the overall shape of the $w(z)$ curve in both parametrizations. 

 Both cases were evaluated together with the CPL parametrization (Eq. \ref{eq:cpl}). We performed a MCMC run by introducing the modified CAMB routines and the new parametrization for $w(z)$ in the COSMOMC code. The total parameter space directly constrained was then {$\mathbf p$}=$\{\Omega_b h^2, \Omega_c h^2, \theta, \tau, A_s, n_s,$ $w_0, w_1 \}$.   
We used the matter power spectrum from \cite{T06}  to account for the SDSS LRG data. The modeling for non linear effects and scale dependent bias was done with a Q-model (\cite{Cole05}) as explained above. Since we are interested in  models with $w(z)<-1$ in order to determine a valid figure of merit, dark energy perturbations were ``turned off'' when solving for CMB and matter power spectrum.

\begin{table*}[h!]
\caption{Table resuming the obtained constraints for a pure CPL model (1), parameterization \ref{wqz} with $\Gamma=1$ (2), $\Gamma=0.85$ (3) and assuming a constant w (4).} 
\centering
\begin{tabular}{c c c c c}
\hline
$\mathbf{Param.}$	& $\mathbf{1}$	& $\mathbf{2} $	 &$\mathbf{3}$	  &$\mathbf{Vanilla + w}$\\
\hline
$        \Ob h^2\,^{\mathrm[a]}$        &  $0.0222\pm0.0006$   &$ 0.0228\pm0.0006$                            &  $0.0228\pm0.006 $           & $0.0228\pm0.0006$   	\\        
$           \Omega_c h^2\,^{\mathrm[b]}$ &  $0.107\pm0.005$   &$0.107\pm0.005 $                     &  $0.107\pm0.005$           & $0.109\pm0.005$     \\
$                 \theta\,^{\mathrm[c]}$ & $0.104\pm0.003 $   &$ 1.041\pm0.002$                      &  $1.041\pm0.002$           & $1.042\pm0.003$	  \\
$                 \tau\,^{\mathrm[d]}$   &  $0.856\pm0.017$   &$ 0.088\pm0.017$                     &  $0.088\pm0.016$           & $0.087\pm0.017$  \\  
$             n_s\,^{\mathrm[e]}$       &    $0.958\pm0.014$  &$0.969\pm0.008$                     &  $0.968\pm0.008$           & $0.967\pm0.014$ \\  
$        log(10^{10} A_s)\,^{\mathrm[f]}$&   $3.05\pm0.04$   &$3.06\pm0.04$                      &  $3.06\pm0.04$            & $3.06\pm0.04$	\\
$                     w_0\,^{\mathrm[g]}$&   $-1.027\pm0.123$   &$ -0.994\pm0.141 $                  &  $ -1.005\pm0.119 $       &  $-0.965\pm0.056$        \\
$                     w_1\,^{\mathrm[g]}$&   $0.278\pm0.408$   &$ 0.036\pm0.452 $                   &  $ 0.186\pm0.404$           &  $0$	\\
\hline
$                 \Omega_Q (\Ol)\,^{\mathrm[h]}$   &  $0.735\pm0.017$   &$ 0.747\pm0.018         $   &   $0.738\pm0.018$           & $0.739\pm0.014$    	\\
$                      Age\,^{\mathrm[i]}       $  &   $13.9\pm0.1$  &$  13.7\pm0.1$              &  $13.8\pm0.1$            & $13.7\pm0.1$	         \\
$                      \Om\,^{\mathrm[j]}$         &   $0.265\pm0.017$    &$ 0.253\pm0.018$                   &  $0.262\pm0.018$            & $0.261\pm0.020$	 \\
$                    \sigma_8\,^{\mathrm[k]}$      &  $0.766\pm0.037$     &$0.780\pm0.037$                 &  $0.776\pm0.037$            & $0.816\pm0.014$     	 \\
$                   z_{re}\,^{\mathrm[l]}$         &  $11.0\pm1.5$    &$11.1\pm1.4$                       &  $11.1\pm1.4$            &  $11.0\pm1.5$	   \\
$                    h\,^{\mathrm[m]} $            &  $0.698\pm0.018$    &$ 0.718\pm0.020$                      &  $0.706\pm0.017$          & $0.713\pm0.015$	 	\\

\hline
\end{tabular}
\begin{list}{}{}
\item$^{\mathrm[a]}$ Normalized baryon matter density times $h^2$, $^{\mathrm[b]}$ normalized CDM density times $h^2$, $^{\mathrm[c]}$ ratio of sound horizon to angular diameter distance, $^{\mathrm[d]}$ optical depth to reionization, $^{\mathrm[e]}$  primordial spectral index, $^{\mathrm[f]}$  primordial scalar power at $k=0.05 \rm Mpc^{-1}$, $^{\mathrm[g]}$  parameters describing the equation of state of dark energy, $^{\mathrm[h]}$  dark energy density, $^{\mathrm[i]}$  age of the universe in Gyr, $^{\mathrm[j]}$  normalized matter density, $^{\mathrm[k]}$  matter density fluctuation amplitude, $^{\mathrm[l]}$ reionization redshift, $^{\mathrm[m]}$ reduced Hubble parameter.

\end{list}
\label{Tablew0w1}
\end{table*}                

\begin{figure}[t!]
\centering
\includegraphics[width=0.45\textwidth]{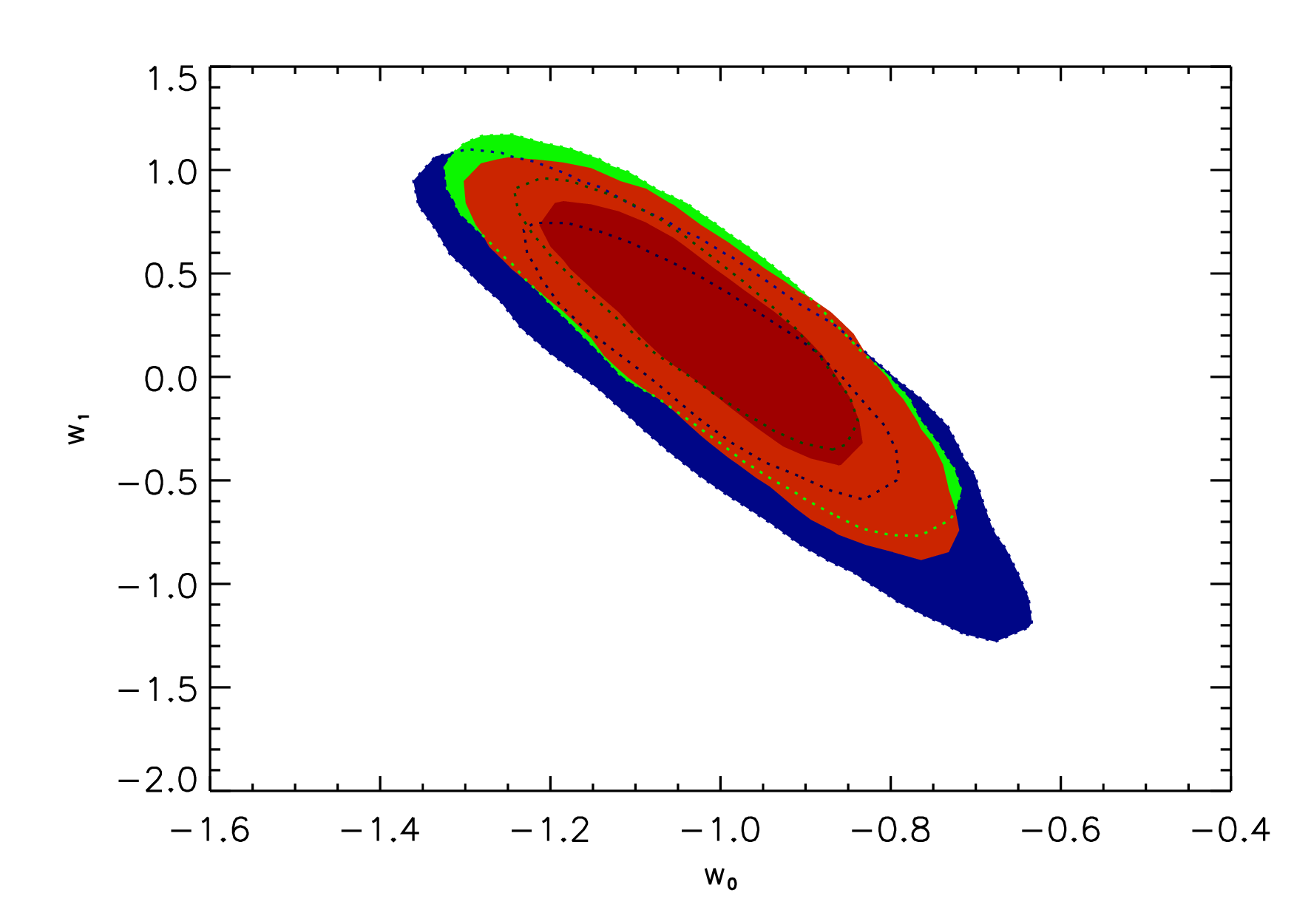}
\caption{2D contours at 68 \% and 95\% confidence level, for the parameters $w_0=1/2(w_++w_-)$ and $w_1=1/2(w_+-w_-)$, setting $z_t=1$. The blue contours correspond to $\Gamma=1$ and the red contours to $\Gamma=0.85$. The green countour shows the same confidence regions using the standard CPL parameterization.}
\label{fig:w0w1}
\end{figure}

The obtained results on all parameters are summarized in Table \ref{Tablew0w1}. All these constraints on individual parameters were obtained by marginalization over other parameters. For comparison reasons, the last column shows the results where $w$ is a constant value (no evolution). The constraints on nearly all parameters remain essentially identical. Of course, the main exception regards parameters that govern the equation of state itself, which have  a bigger uncertainty compared to a constant $w$. Considering the parametrization of a smooth transition with either $\Gamma=1$ or $\Gamma=0.85$ results in some small changes in the preferred values and the uncertainties, which seem to be smaller when $\Gamma=0.85$. 
The 2D marginalized contours at 68\% and 95 \% confidence level are presented  in Fig. \ref{fig:w0w1}.
Again, the slightly different 
values of the  $\Gamma$ parameter have an effect on the derived constraints on $w_0-w_1$, although it is rather small. Finally, we notice that using the parametrization given by Eq. \ref{wqz} with $\Gamma=0.85$ provides very similar constraints 
to a pure CPL parametrization. 

The Dark Energy Task Force report (\cite{DETF})  introduced a numerical quantity to determine the capacity of future surveys in constraining dark energy models. This numerical quantity is referred to as the figure of merit (f.o.m.) for a given experiment or combination of experiments and is defined as the reciprocal of the area of the marginalized 2D region contour at 95 \% confidence in the $w_0-w_1$ parameter space. 
In terms of the f.o.m. for the present day data, the obtained values are summarized in the following table  

\begin{table}[h!]
\centering
\begin{tabular}{c c c c}
Model & CPL & $\Gamma=1$ & $\Gamma=0.85$ \\
\hline
f.o.m. & $2.03$ & $1.54$ & $2.11$ \\
\end{tabular}
\end{table}

These values can thus be used to compare the improvement of future experiments in constraining dark energy evolution on Hubble timescales. For more rapid transitions, one has to consider higher values of $\Gamma$, which is done in the next section.                    

\subsection{Joint constraints}

In this section we  examine what constraints can be derived when a rapid transition in the equation of state of dark energy ($\Gamma=5$) is considered, using the full data sets 
presented above. 
Only models with $w(z)>-1$ were constrained by including DE perturbations when solving the total perturbations equations with CAMB. For each case we ran one long chain, using the Rafetery-Lewis diagnostic to check for convergence. As for the LRG SDSS data, we used the data on the matter power spectrum as implemented in COSMOMC with a Q-model to correct for non-linear effects.
 Finally, we restricted ourselves to a flat Universe ($\Omega_k=0$) for simplicity reasons.  

As seen above, for transitions to $w_-=-1$, the matter correlation function 
shows a different sensitivity to the initial value $w_+$. We
considered then three distinct transitions by setting $w_-=-1$ and three low values
 for $w_+$ ($0, -0.1$ and $-0.2$). In each case the objective is to get 
constraints on the transition epoch ($a_t$),  the value $a_t= 0$ 
corresponding to the standard
concordance model.   As $w_+$ approaches $0$ the contribution of dark energy at high redshifts remains non vanishing, allowing us to investigate its 
possible influence at earlier epochs.  

\subsubsection{$\bullet$ $w_-=-1, w_+=-0.2$}

 \begin{table*}[h!]
\caption{Table resuming the obtained constraints different combinations of data and setting $w_+=-0.2$ and $w_-=-1$.}
\centering
\begin{tabular}{c c c c c}
\hline
$\mathbf{Param.}$		&$\mathbf{CMB+SN~Ia} $	 &$\mathbf{CMB+P(k)}$			&$\mathbf{CMB+P(k)+SN~Ia}$\\
\hline
$        \Ob h^2$              &$ 0.0225\pm0.0007$                            &   $0.0224\pm0.0006 $           & $0.0224\pm0.0006$   	\\        
$           \Omega_c h^2$      &$0.107\pm0.007 $                     &  $0.104\pm0.005$           & $0.105\pm0.005$     \\
$                 \theta$      &$ 1.040\pm0.003$                      &  $1.040\pm0.003$           & $1.039\pm0.003$	  \\
$                 \tau$        &$ 0.087\pm0.019$                     &  $0.088\pm0.017$           & $0.085\pm0.016$  \\  
$             n_s$             &$0.969\pm0.018$                     &  $0.967\pm0.016$           & $0.968\pm0.017$ \\  
$        log(10^{10} A_s)$      &$3.05\pm0.04$                      &  $3.04\pm0.04$            & $3.04\pm0.04$	\\
$                     a_t$      &$ 0.31_{-0.31}^{+0.09} $                  &  $ 0.25_{-0.25}^{+0.08} $       &  $ 0.29_{-0.29}^{+0.09}$        \\

\hline
$                 \Omega_Q (\Ol)$      &$ 0.718\pm0.023         $   &  $0.743\pm0.020$           & $0.733\pm0.017$    	\\
$                      Age       $      &$  14\pm0.2$               &  $13.9\pm0.2$            & $14.0\pm0.2$	         \\
$                      \Om$      &$ 0.282\pm0.023$                  &  $0.257\pm0.020$            & $0.267\pm0.017$	 \\
$                    \sigma_8$      &$0.705\pm0.086$                &  $0.717\pm0.063$            & $0.707\pm0.075$     	 \\
$                   z_{re}$      &$11.1\pm1.6$                      &  $11.1\pm1.5$            &  $11.0\pm1.4$	   \\
$                    h $      &$ 0.678\pm0.028$                     &  $0.704\pm0.025$          & $0.691\pm0.024$	 	\\

\hline
\end{tabular}
\label{Table_wi02wf1}
\end{table*}

\begin{figure}[h!]
\centering
\includegraphics[width=0.45\textwidth]{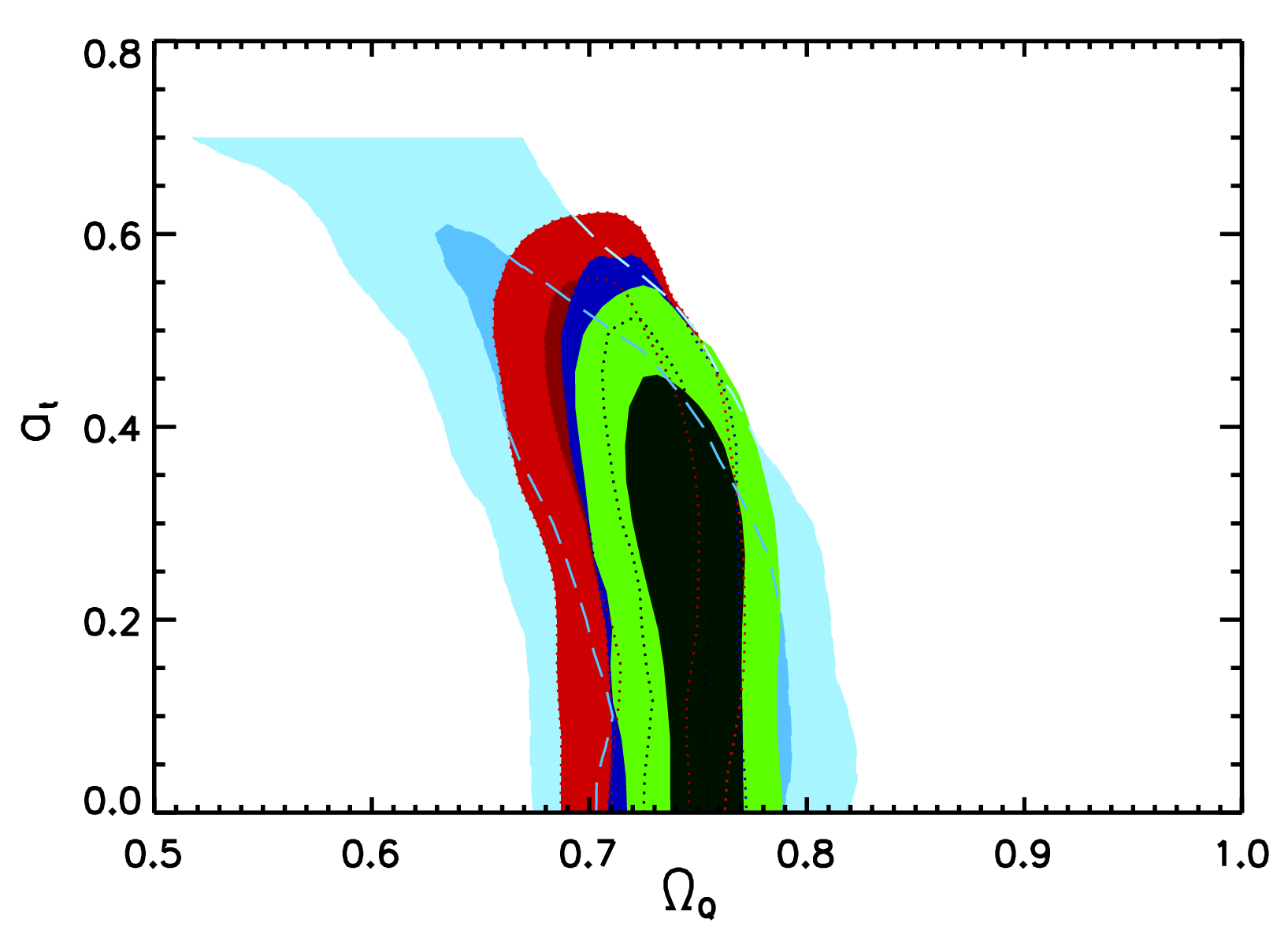}\\
\caption{Confidence regions at 68 \% and 95 \% level for the transition epoch $a_t$ and the density of the quintessence field $\Omega_Q$, considering $w_+=-0.2$ and $w_-=-1$ and using different combinations of data:CMB (\emph{light blue}), CMB+ SN~Ia (\emph{red}), CMB+P(k)(\emph{green}) and CMB+SN~Ia+P(k)(\emph{blue}). }
\label{fig:Ol_at_1}
\end{figure}

\noindent Four different data combinations were used to constrain this model, CMB, CMB+SN~Ia, CMB+P(k) and finally CMB+P(k)+ SN~Ia. The results of the mean posterior values and 68\% confidence intervals on each parameter are summarized in Table \ref{Table_wi02wf1}. The posterior distribution in some parameter is clearly
non-Gaussian, so in some cases the 1$\sigma$ error bars are quoted using asymmetrical limits.
Fig. \ref{fig:Ol_at_1} shows the  contours in the space of $\Omega_Q-a_t$, on which  
the effect of 
 the different data combinations on these two parameters can be appreciated. This figure shows that the combination of the WMAP5 data on the CMB with supernovae already constrains the possible transition to be at $a_t<0.6$ at the two sigma level (95\% confidence region), or in redshift space $z_t>0.7$. When the matter power spectrum data is added, some small reduction on the allowed region in the transition parameter $a_t$ is observed.
 Surprisingly,  the combination CMB+P(k) without supernovae
leads to a better constraint on the transition epoch. This result is not intuitive, since as demonstrated in the previous section, one obtains only negligible differences in the matter correlation function shape by changing the transition epoch. The cause for this reduction of the allowed transition redshift is actually due to the shift in the preferred $\Om $, as the power spectrum data  tend to prefer lower values of $\Omega_m$.

\begin{figure}[h!]
\centering

\includegraphics[width=0.45\textwidth]{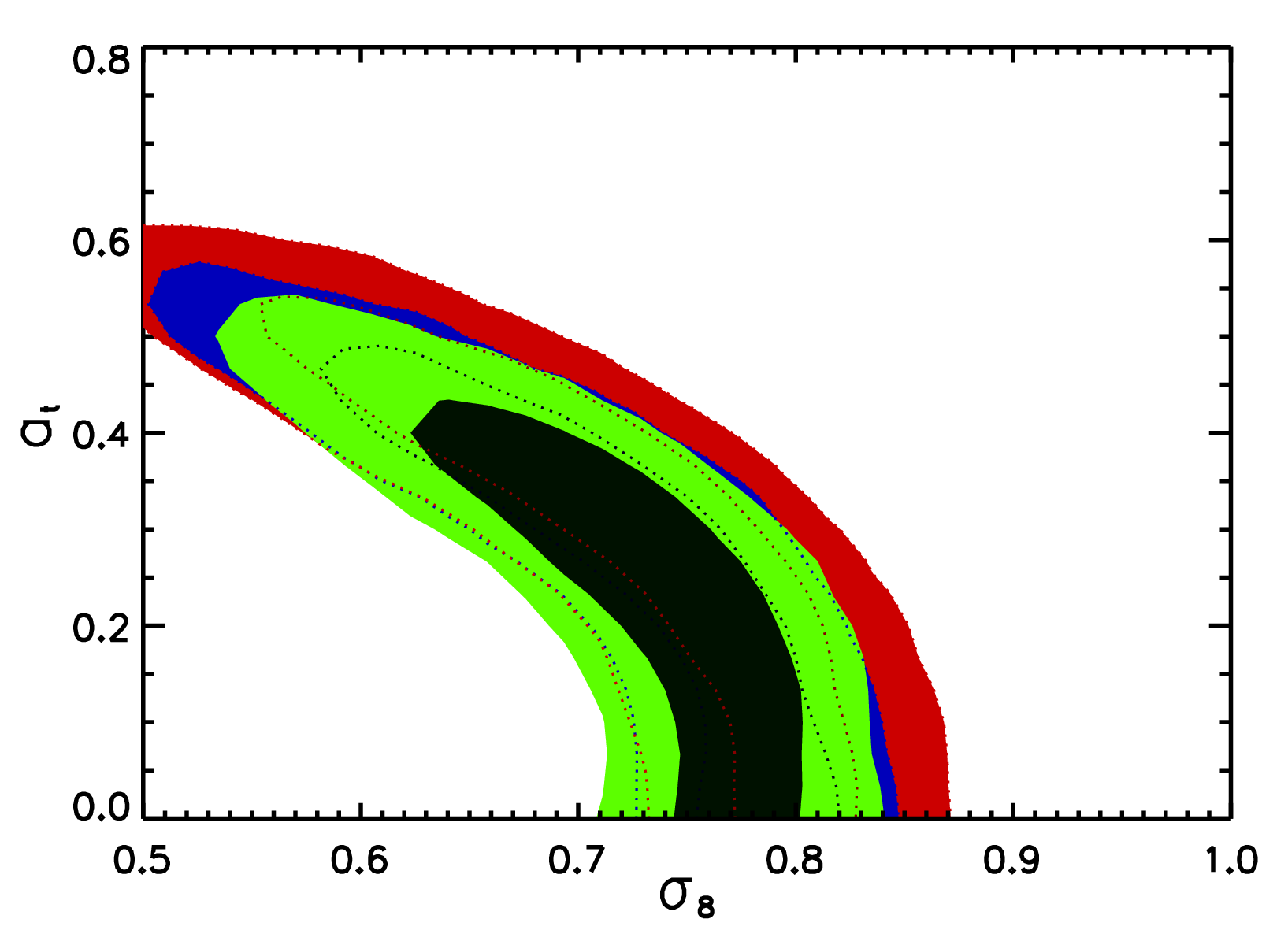}\\
\caption{Confidence regions at 68 \% and 95 \% level for the transition epoch $a_t$ and the amplitude of matter perturbations $\sigma_8$, considering $w_+=-0.2$ and $w_-=-1$ and using different combinations of data: CMB+ SN~Ia (\emph{red}), CMB+P(k)(\emph{green}) and CMB+SN~Ia+P(k)(\emph{blue}).}
\label{fig:sigma_at_1}
\end{figure}

Another interesting factor of this dark energy modeling is the effect 
that transitions at intermediate redshifts have in the allowed matter fluctuation amplitude - the $\sigma_8$ parameter. Figure \ref{fig:sigma_at_1} 
presents a peculiar degeneracy between these parameters. This figure shows that for transitions occurring between $a_t \approx 0.2$ ($z_t=4$) and $a_t\approx 0.6$ ($z_t=0.6$), the allowed matter amplitude $\sigma_8$ decreases linearly, meaning that transitions which occur at lower redshifts necessarily yield  a lower value of $\sigma_8$. This is because a value of $w$ close to $0$ has the effect of inhibiting the growth of matter perturbations, as shown in \cite{Douspis2003}. 
So, in the presence of a transition from a value of $w$ close to 0, the
 overall amplitude of fluctuations (for a given CMB amplitude) is smaller  at low redshifts. In principle the amplitude of matter fluctuations can be estimated from local cluster abundances. However, in practice the lack of reliable  calibration of the mass temperature limits the use of this constraint (\cite{Pierpaoli}; \cite{ATM}).

\subsubsection{$w_-=-1, w_+=-0.1$} 

\begin{figure}[t!]
\centering
\includegraphics[width=0.45\textwidth]{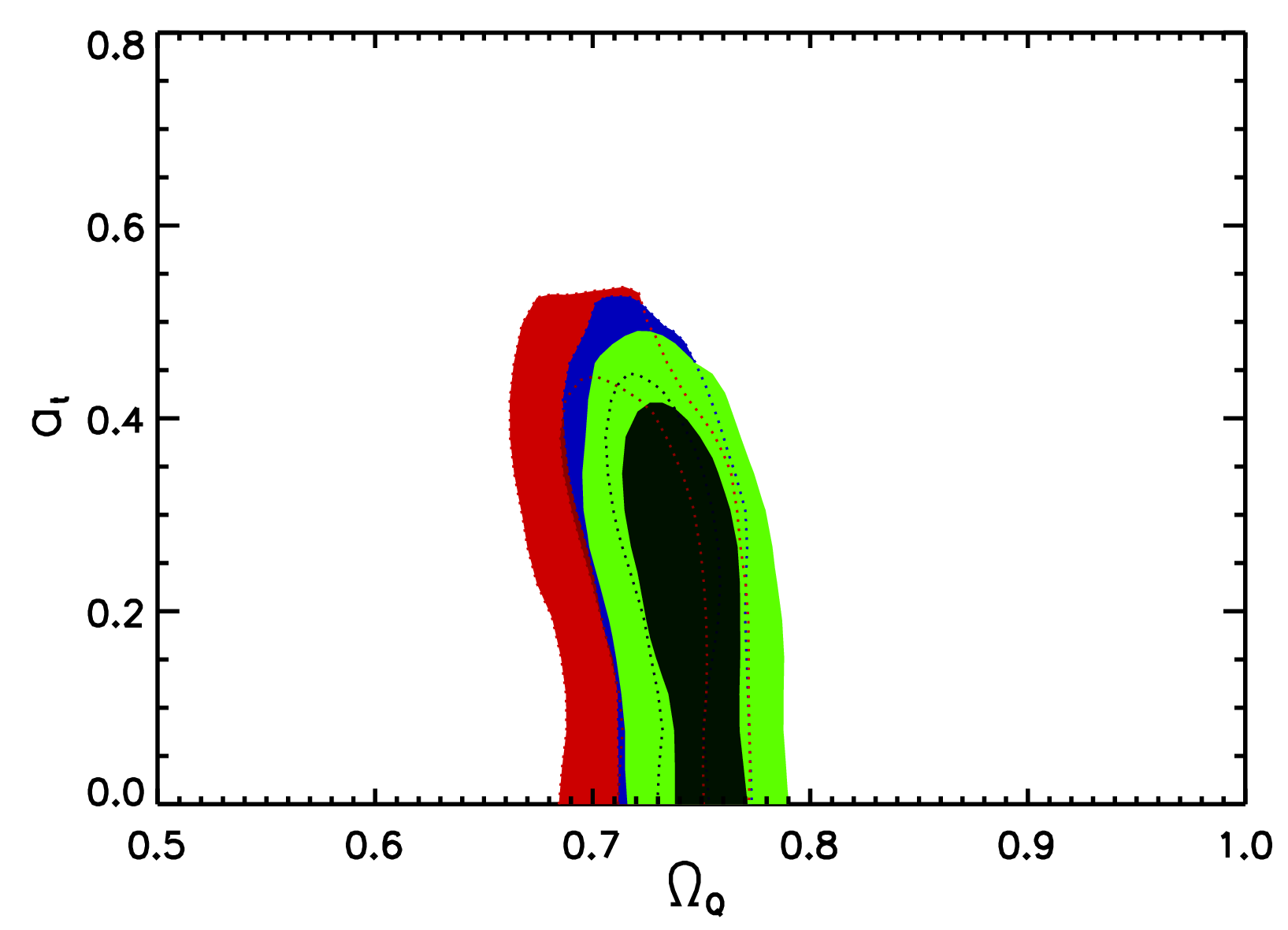}\\
\caption{Confidence regions at 68 \% and 95 \% level for the transition epoch $a_t$ and the density of the quintessence field $\Omega_Q$, considering $w_+=-0.1$ and $w_-=-1$ and using different combinations of data: CMB+SN~Ia (\emph{red}), CMB+P(k)(\emph{green}) and CMB+SN~Ia+P(k)(\emph{blue}).}
\label{fig:Ol_at_wi01_wf1}
\end{figure}

We have seen above that 
for a value of $w_+$ closer to 0 important changes in the predicted matter correlation function arise, which can improve the quality of the fit of the SDSS LRG data. We now check whether these models, which comprise transitions at low redshifts, may accommodate other observational data. So, we determine the same combined constraints as done previously, but setting $w_+=-0.1$. 
The obtained results for the other data combinations are summarized in Table \ref{Table_wi01wf1} and in Fig. \ref{fig:Ol_at_wi01_wf1}, which shows the marginalized region contours in the $\Omega_Q - a_t$ parameter space. 
For these models the allowed transition redshift space is reduced: 
models with $\Omega_Q\simeq0.75$ ($\Omega_m\simeq0.25$) and $a_t>0.5$ ($z_t<1$) are excluded at the two sigma level. This
  means that for these models the CMB data impose a stronger constraint, and the best fit model in middle panel of Fig. \ref{fig:xi} with $z_t=0.5$ becomes ruled out by the CMB.           

  \begin{table*}[h!]
\caption{Table resuming the obtained constraints using different combinations of data and setting $w_+= -0.1$ and $w_-=-1$.}
\centering
\begin{tabular}{c c c c c}
\hline
$\mathbf{Param.}$		&$\mathbf{CMB+SN~Ia} $	 &$\mathbf{CMB+P(k)}$			&$\mathbf{CMB+P(k)+SN~Ia}$\\
\hline
$        \Ob h^2$              &$ 0.0224\pm0.0007$                            &   $0.0224\pm0.0006 $           & $0.0224\pm0.0007$   	\\        
$           \Omega_c h^2$      &$0.109\pm0.006 $                     &  $0.105\pm0.005$           & $0.106\pm0.005$     \\
$                 \theta$      &$ 1.040\pm0.003$                      &  $1.040\pm0.003$           & $1.039\pm0.003$	  \\
$                 \tau$        &$ 0.089\pm0.017$                     &  $0.089\pm0.017$           & $0.088\pm0.018$  \\  
$             n_s$             &$0.970\pm0.022$                     &  $0.971\pm0.019$           & $0.972\pm0.020$ \\  
$        log(10^{10} A_s)$      &$3.06\pm0.04$                      &  $3.05\pm0.04$            & $3.04\pm0.04$	\\
$                     a_t$      &$ 0.25_{-0.31}^{+0.08} $                  &  $ 0.23_{-0.23}^{+0.07} $       &  $ 0.26\pm_{-0.26}^{+0.07}$        \\

\hline
$                 \Omega_Q (\Ol)$      &$ 0.719\pm0.023         $   &  $0.743\pm0.020$           & $0.733\pm0.017$    	\\
$                      Age       $      &$  14\pm0.2$               &  $13.9\pm0.2$            & $14.0\pm0.2$	         \\
$                      \Om$      &$ 0.281\pm0.023$                  &  $0.257\pm0.020$            & $0.267\pm0.017$	 \\
$                    \sigma_8$      &$0.719\pm0.086$                &  $0.708\pm0.074$            & $0.696\pm0.083$     	 \\
$                   z_{re}$      &$11.4\pm1.5$                      &  $11.1\pm1.5$            &  $11.2\pm1.6$	   \\
$                    h $      &$ 0.686\pm0.026$                     &  $0.704\pm0.024$          & $0.693\pm0.023$	 	\\

\hline
\end{tabular}

\label{Table_wi01wf1}
\end{table*}    
\subsubsection{$w_-=-1, w_+=0$} 

\begin{figure}[h!]
\centering
\includegraphics[width=0.45\textwidth]{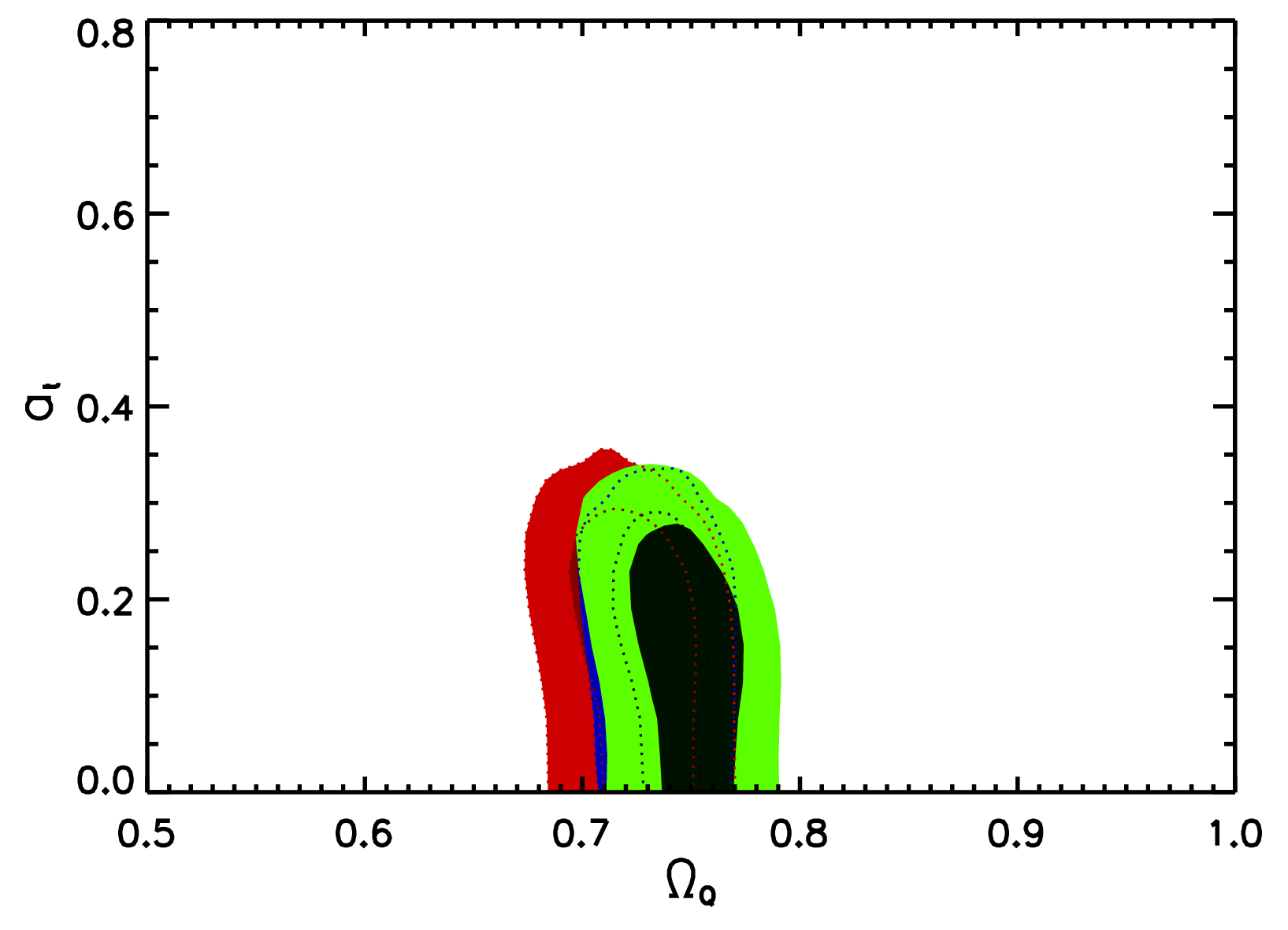}\\
\caption{Confidence regions at 68 \% and 95 \% level for the transition epoch $a_t$ and the density of the quintessence field $\Omega_Q$, considering $w_+= 0.$ and $w_-=-1$ and using different combinations of data: CMB+SN~Ia (\emph{red}), CMB+P(k)(\emph{green}) and CMB+ SN~Ia+P(k)(\emph{blue}).}
\label{fig:Ol_at_wi0_wf1}
\end{figure}

Finally we consider the case where the dark energy  has a vanishing equation of state parameter in the early Universe. This case is phenomenologically interesting as dark energy then evolves as pressureless matter, although it cannot be accounted for in standard quintessence models like the inverse power law or SUGRA potentials (\cite{Douspis2008}). In Table \ref{Table_wi0wf1} and Fig. \ref{fig:Ol_at_wi0_wf1} we present 
the same information as above for the constraints on this model. The result is somehow expected from what we saw when we passed from $w_+-0.2$ to $w_+=-0.1$. While the constraints on $\Om$ ($\Omega_Q$) remain almost unchanged, the possible transition epoch is further pushed towards higher redshifts, with a value $z_t>2$ at 95\% confidence level. 
 \begin{table*}[h!]
\caption{Table resuming the constraints using different combinations of data and setting $w_+=0$ and $w_-=-1$. }
\centering
\begin{tabular}{c c c c c}
\hline
$\mathbf{Param.}$		&$\mathbf{CMB+SN~Ia} $	 &$\mathbf{CMB+P(k)}$			&$\mathbf{CMB+P(k)+SN~Ia}$\\
\hline
$        \Ob h^2$              &$ 0.0223\pm0.0006$                 &   $0.0223\pm0.0006 $        & $0.0223\pm0.0006$   	\\        
$           \Omega_c h^2$      &$0.112\pm0.005 $                   &  $0.108\pm0.004$           & $0.110\pm0.004$     \\
$                 \theta$      &$ 1.038\pm0.003$                   &  $1.038\pm0.003$           & $1.038\pm0.003$	  \\
$                 \tau$        &$ 0.086\pm0.016$                  &  $0.088\pm0.017$            & $0.087\pm0.017$  \\  
$             n_s$             &$0.965\pm0.016$                    &  $0.968\pm0.016$           & $0.965\pm0.015$ \\  
$        log(10^{10} A_s)$     &$3.06\pm0.04$                     &  $3.05\pm0.04$              & $3.05\pm0.04$	\\
$                     a_t$     &$ 0.16_{-0.16}^{+0.05} $          &  $ 0.15_{-0.15}^{+0.05} $     &  $ 0.16_{-0.16}^{+0.05}$        \\

\hline
$                 \Omega_Q (\Ol)$      &$ 0.723\pm0.021         $   &  $0.746\pm0.020$           & $0.737\pm0.016$    	\\
$                      Age       $      &$  13.8\pm0.1$               &  $13.8\pm0.1$            & $14.0\pm0.1$	         \\
$                      \Om$      &$ 0.277\pm0.021$                  &  $0.254\pm0.020$            & $0.263\pm0.016$	 \\
$                    \sigma_8$      &$0.759\pm0.052$                &  $0.740\pm0.050$            & $0.746\pm0.048$     	 \\
$                   z_{re}$      &$11.1\pm1.4$                      &  $11.1\pm1.5$            &  $11.1\pm1.5$	   \\
$                    h $      &$ 0.698\pm0.018$                     &  $0.717\pm0.019$          & $0.707\pm0.015$	 	\\

\hline
\end{tabular}

\label{Table_wi0wf1}
\end{table*}    
Again, this constraint comes mainly from the combination of the CMB observations with the other probes.

From the three considered cases, 
the models with late transitions $z_t<1$ seem to be excluded since they cause the angular distance to the CMB to change significantly which disagrees with the position of the acoustic peak in the temperature power spectrum. This finding does not allow for lower values of the transition redshift when combined with the strong constraint on $\Om$ from the SDSS LRG distribution at low redshifts and the SN~Ia data.

\subsubsection{Relaxing the assumption of $w_-=-1$}

We have seen the effects of considering transitions that have a final value $w_-$ compatible with a cosmological constant, i.e. $w\sim -1$ at low redshifts and in the future. This assumption was based on the theoretical prior that a final $w=-1$ is easier to understand in existing quintessence models. Nevertheless, one can relax this assumption on a purely phenomenological basis, and check what happens to the constraints derived above when $w_-$ is left as a free parameter. We have performed this analysis, taking the same three values for $w_+$ as before 
and using the full combination of data CMB+P(k)+SN~Ia. 
  \begin{table*}[h!]
\caption{Table resuming the obtained constraints using the full combination of data (CMB+P(k)+SN~Ia) and considering $w_-$ as a free parameter.}
\centering
\begin{tabular}{c c c c c}
\hline
$\mathbf{Param.}$		&$\mathbf{w_+=0} $	 &$\mathbf{w_+=-0.1}$			&$\mathbf{w_+=-0.2}$\\
\hline
$        \Ob h^2$              &$ 0.0222\pm0.0006$                            &   $0.0223\pm0.0006 $           & $0.0223\pm0.0006$   	\\        
$           \Omega_c h^2$      &$0.106\pm0.005 $                     &  $0.106\pm0.005$           & $0.107\pm0.005$     \\
$                 \theta$      &$ 1.039\pm0.003$                      &  $1.039\pm0.003$           & $1.042\pm0.003$	  \\
$                 \tau$        &$ 0.085\pm0.017$                     &  $0.085\pm0.016$           & $0.085\pm0.017$  \\  
$             n_s$             &$0.969\pm0.008$                     &  $0.960\pm0.013$           & $0.960\pm0.014$ \\  
$        log(10^{10} A_s)$      &$3.05\pm0.04$                      &  $3.05\pm0.04$            & $3.05\pm0.04$	\\
$                     a_t$      &$ 0.047_{-0.047}^{+0.013} $        &  $ 0.071_{-0.071}^{+0.018} $  &  $ 0.100_{-0.100}^{+0.026}$        \\
$                     w_-$      &$ -0.954\pm0.051 $                  &  $ -0.951\pm0.058 $       &  $-0.965\pm0.056$        \\
\hline
$                 \Omega_Q (\Ol)$      &$ 0.742\pm0.015         $   &   $0.742\pm0.015$           & $0.740\pm0.014$    	\\
$                      Age       $      &$  13.8\pm0.1$              &  $13.8\pm0.1$            & $13.8\pm0.1$	         \\
$                      \Om$      &$ 0.258\pm0.015$                   &  $0.258\pm0.015$            & $0.260\pm0.015$	 \\
$                    \sigma_8$      &$0.763\pm0.036$                 &  $0.761\pm0.038$            & $0.764\pm0.037$     	 \\
$                   z_{re}$      &$11.0\pm1.5$                       &  $11.0\pm1.4$            &  $10.9\pm1.5$	   \\
$                    h $      &$ 0.705\pm0.016$                      &  $0.705\pm0.016$          & $0.704\pm0.016$	 	\\

\hline
\end{tabular}

\label{Table_wi02wf}
\end{table*}  
 
\begin{figure}[h!]
\centering
\includegraphics[width=0.45\textwidth]{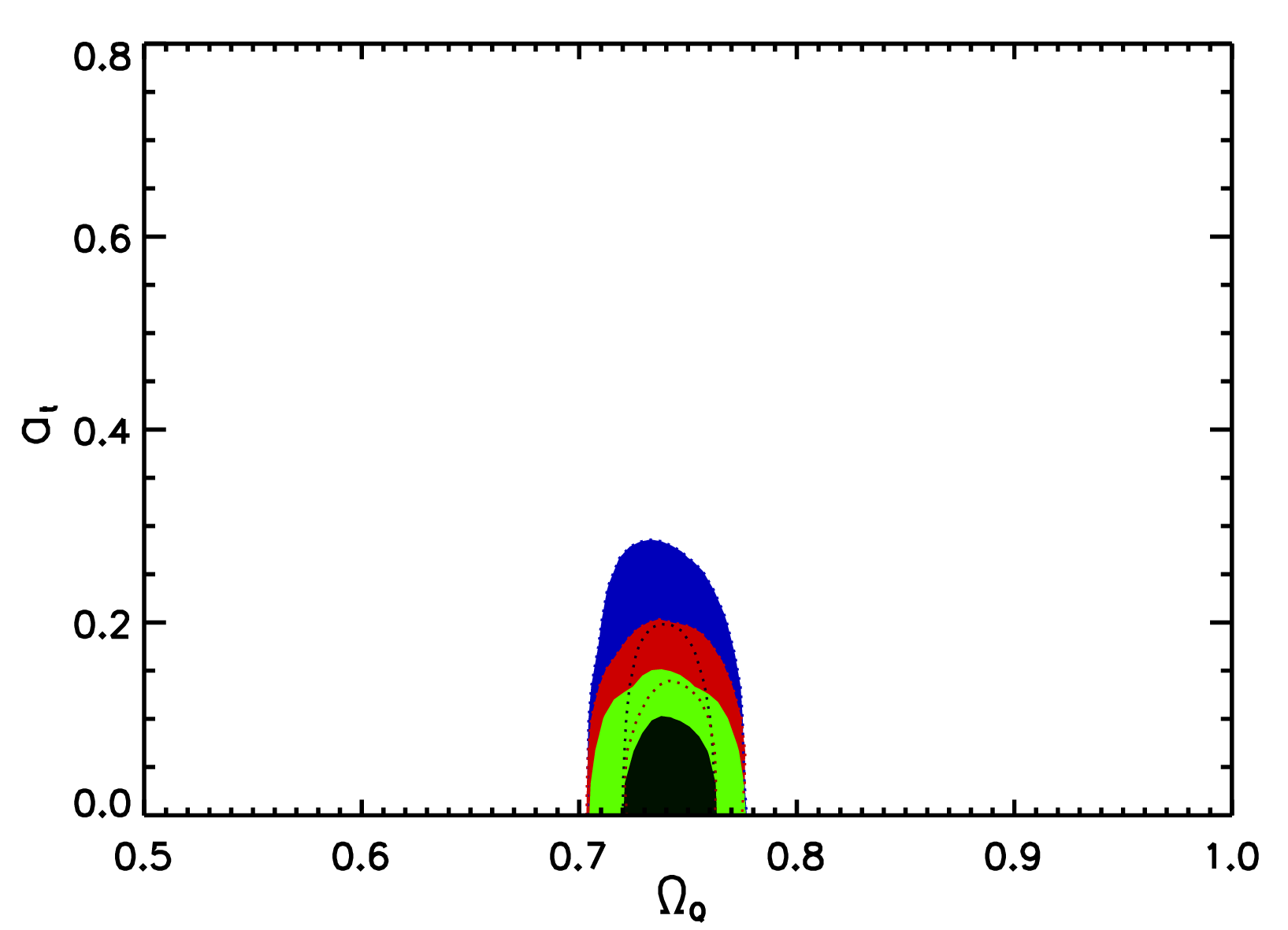}\\
\caption{Confidence regions at 68 \% and 95 \% level for the transition epoch $a_t$ and the density of the quintessence field $\Omega_Q$,  $w_-$ being an additional  parameter  on which the likelihood is marginalized  and using different values for $w_+$: 0 (\emph{green}), -0.1 (\emph{red}) and -0.2 (\emph{blue}) using CMB+SN~Ia+P(k).}
\label{fig:Ol_at_2}
\end{figure}

\begin{figure}[h!]
\centering
\includegraphics[width=0.45\textwidth]{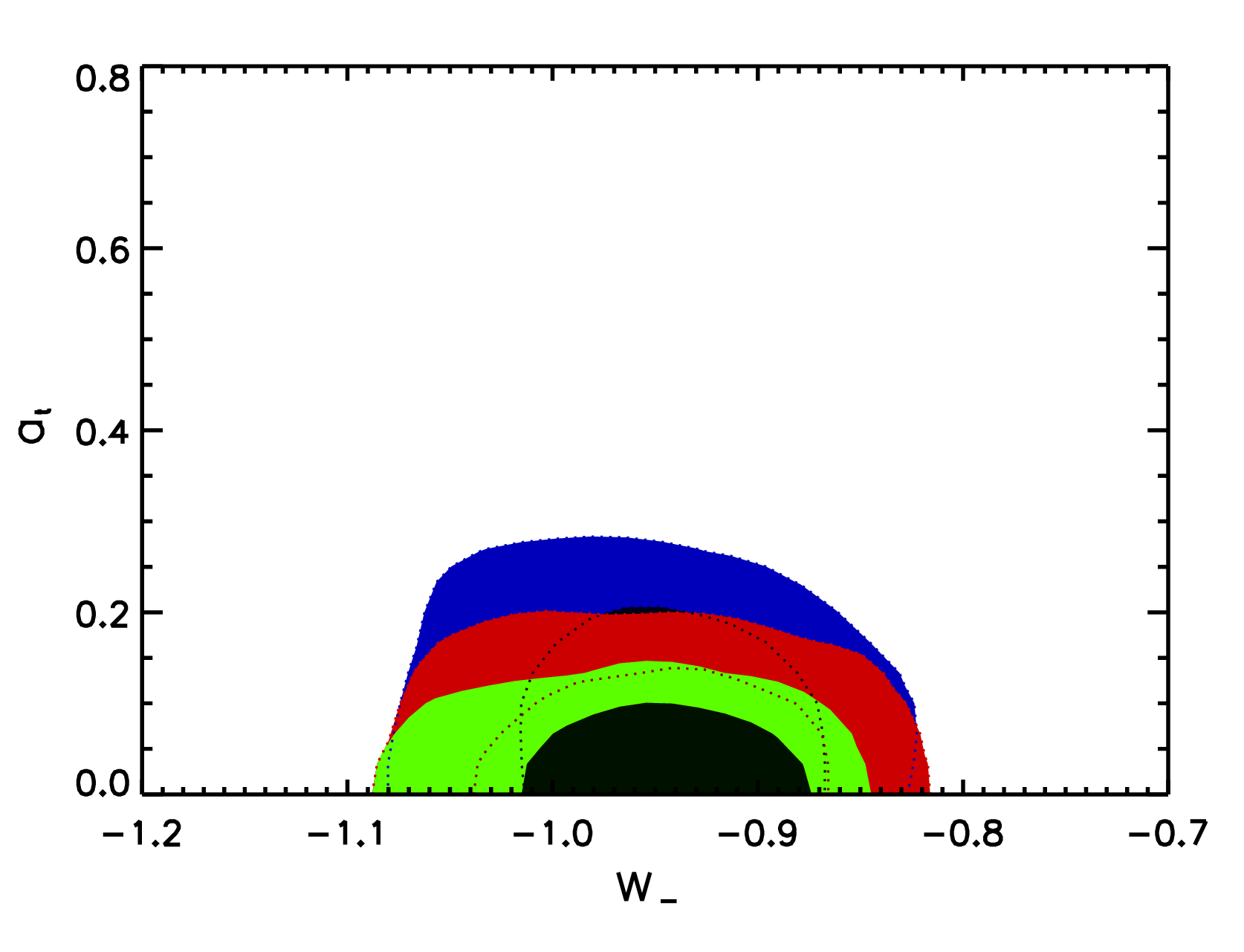}\\
\caption{Confidence regions at 68 \% and 95 \% level for the transition epoch $a_t$ and $w_-$, marginalizing on $\Omega_Q$ with different values for $w_+$: 0 (\emph{green}), -0.1 (\emph{red}) and -0.2 (\emph{blue}).}
\label{fig:Ol_at_2b}
\end{figure}

The results on the three chains are summarized in Table \ref{Table_wi02wf}. 
The main constraints on the transition parameters are summarized  in Fig. \ref{fig:Ol_at_2} and \ref{fig:Ol_at_2b} which show the effect of leaving $w_-$ as a free parameter in the transition epoch defined by $a_t$. 
 The effect is drastic, pushing the possible transition to much higher redshifts while at the same time the preferred value for $w_-$ becomes $w_-=-0.956$, with $w_-=-1$ just compatible at one sigma. That the constraint put on $a_t$ for the different values of $w_+$ becomes much more severe ($z_t>4$ at two sigma, in the best case) may seem surprising since the obtained value is not so far from the fiducial $w=-1$ considered above. As a check,
we reevaluated our constraints setting  $w_-=-0.95$ and found identical constraints to those marginalized over $w$.     

It seems that by letting the present day value for the equation of state be a little higher than $-1$, current data tend to strongly prefer the case of no transition, rejecting any acceptable transition to higher redshifts. This agrees with the constraints obtained when we considered for a constant value of $w$ previously, which is a good cross-check of this analysis. That the value $w=-0.95$ is systematically preferred is somehow interesting, although with the current constraint precision, the $\Lambda$CDM is obviously compatible with all derived constraints.

\section{Discussion and conclusions}
\label{secconc}

We have investigated quintessence  models which undergo a 
transition in the equation of state parameter $w$ of their dark energy 
component in
the light of latest cosmological data. 
We concentrated on a class of models in which the duration of the transition  is much
shorter than the Hubble time, for which we have  previously shown that the supernovae Hubble diagram alone  provides, somewhat surprisingly, poor  constraints. A
 major difference to previous studies of varying dark energy models is that we fully took into account the data on the matter power spectrum rather than using a reduced geometrical factor $A$ and used a class of models in which the dark energy component 
at high redshift does not cross the matter regime, i.e. $w$ remains below $0$.
We concentrated on  models in which the high redshift equation of state parameter approaches zero (specifically $w = -0.2$, $w = -0.1$, $w = 0.$), without marginalizing because of the sensitivity to $w$ when it approaches zero. The first 
noticeable result is the stability of cosmological constraints obtained for the fundamental parameters: all parameters that our five tables have in common 
are nearly unchanged in comparison with those obtained for a
constant $w$ (\cite{FBZ}).  In this respect we noticed that a value of $w =  -0.95\pm 0.05$ remains preferred at the one sigma level. The amplitude of matter fluctuations reveals some sensitivity to dark energy dynamics, as the $\sigma_8$ parmeter is lower in the presence of dark energy with an equation of state parameter $w>-1$, a result consistent with previous findings for constant $w$ (\cite{Douspis2008}). A second conclusion is that a transition is generally severely rejected when combinations of data are used, although a transition would be (weakly) preferred by a single set of data like supernovae or large scale structure. We notice however that the 
redshift at which the possible transition is rejected delicately depends on 
the assumed hypothesis: with a present-day $w$ set to $\sim -1$ and $w_+$ set to $-0.2$
we could formally reject at $2\sigma$ a transition below $z \sim 1.1$, while
 when we marginalize on  $w_-$ and set $w_+ $ to $0$, the transition 
is allowed only at redshifts greater than $\sim 10$. 
 Very recently, new data and analysis have appeared in the literature on SN~Ia (\cite{Freedman2009}; \cite{Kessler2009}; \cite{Lampeitl2009}) and power spectrum (\cite{Percival2009}; \cite{Reid2009}). It is therefore legitimate to wonder whether our conclusions can be appreciably modified with this new information. However, there are some reasons that this is not anticipated. A comparison of the columns in Tables \ref {Table_wi02wf1},\ref{Table_wi01wf1} and \ref{Table_wi0wf1} reveals that our results are not sensitive to different data combinations: using the three data sets (CMB, $P(k)$, SN~Ia) or  using only two (either CMB and $P(k)$ or CMB and  SN~Ia) leads to nearly identical  constraints, differences in preferred parameters being at most 1.5 $\sigma$, less than the worse systematic difference discussed below. Since preferred cosmological parameters from most recent analysis are entirely consistent with our previous analysis (\cite{FBZ}), it is reasonable to assume that using the new data would make virtually no difference.
Our general conclusion is, maybe not surprisingly, that the concordance model is doing well and that possible variation in dark energy is unlikely. 

A final word of caution though: In the present analysis we used the constraint from the power spectrum of LRG 
(\cite{T06}). However, Ferramacho et al. (2009) noticed that the use of the correlation function could lead to preferred values which differ up to $2\sigma$ from the ones obtained with the power spectrum. Therefore, our $2\sigma$ ranges can be regarded as indicative of possible systematic effects. We have no certainty however that this systematic difference could not be larger for varying dark energy models. 
The difference in the clustering properties between red and blue galaxies
is also a subject that calls for caution in the detailed interpretation of data
(\cite{SanchezCole}; \cite{PercivalDR5}), even if recent analysis tends to weaken earlier discrepancies (\cite{Percival2009}).
Improvements of constraints on dark energy from future experiments would have to
control in an extremely accurate way  their systematics to infer robust
constraints on the variation of dark energy. 


\begin{acknowledgements}
       We acknowledge useful comments from J.-M. Virey. L.Ferramacho acknowledges financial support provided by Funda\c c\~ao para a Ci\^encia e Tecnologia (FCT, Portugal) under scholarship SFRH/BD/16416/2004 and project grant PTDC/FIS/66825/2006.  
\end{acknowledgements}



\begin{thebibliography}{99}
\bibitem[Albrecht et al. 2006]{DETF} Albrecht, A., et al.\ 
2006, arXiv:astro-ph/0609591 
\bibitem[Armendariz-Picon et al. 2000]{Kess} 
Armendariz-Picon, C., Mukhanov, V., 
\& Steinhardt, P.~J.\ 2000, Physical Review Letters, 85, 4438 
\bibitem[Armendariz-Picon et al. 2001]{Kess2} 
Armendariz-Picon, C., Mukhanov, V., 
\& Steinhardt, P.~J.\ 2001, \prd, 63, 103510 
\bibitem[Astier  et al.2006]{Astier2006} P. Astier  et al. \ 2006,  A\&A, 447, 31A
\bibitem[Avelino et al. 2009]{Avelino2009} Avelino, P.~P.,
Trindade, A.~M.~M., \& Viana, P.~T.~P.\ 2009, \prd, 80, 067302 
\bibitem[Bamba et al. 2009]{crossing2} Bamba, K., Geng, C.-Q., 
Nojiri, S., \& Odintsov, S.~D.\ 2009, \prd, 79, 083014 
\bibitem[Bassett et al. 2002]{2002MNRAS.336.1217B} Bassett, B.~A., Kunz,
M., Silk, J., \& Ungarelli, C.\ 2002, \mnras, 336, 1217
\bibitem[Blanchard 1984]{Blanchard84} Blanchard, A.\ 1984, \aap, 132, 359 
\bibitem[Blanchard 2010]{Blanchard10} Blanchard, A.\ 2010, \aapr, to be published.  
\bibitem[Blanchard 
\& Douspis 2005]{ATM} Blanchard, A., \& Douspis, M.\ 2005, \aap, 436, 411 
\bibitem[Bouhmadi-L{\'o}pez 
\& Ferrera 2008]{wcross2} Bouhmadi-L{\'o}pez, M., \& Ferrera, A.\ 2008, Journal of Cosmology and Astro-Particle Physics, 10, 11 
\bibitem[Buchert 2008]{Buchert2008} Buchert, T.\ 2008, General 
Relativity and Gravitation, 40, 467 
\bibitem[Caldwell 2002]{Caldwell2002} Caldwell, R.~R.\ 2002, 
Physics Letters B, 545, 23 
\bibitem[Carroll et al. 2003]{wmoins1} Carroll, S.~M., 
Hoffman, M., \& Trodden, M.\ 2003, \prd, 68, 023509 
\bibitem[Chevalier \& Polarski 2001]{CPL1} Chevalier, M.,\& Polarski, D. \ 2001, IJMPD, 10, 213C 
\bibitem[Cognola et al. 2006]{Cognola} Cognola, G.; Elizalde, E.; Nojiri, S. et al.\ 2008, \prd, 77, 046009
\bibitem[Cole et al. 2005]{Cole05} Cole, S., et al.\ 2005, 
\mnras, 362, 505 
\bibitem[Corasaniti et al. 2004]{Corasantini2004} Corasaniti, P.~S., 
Kunz, M., Parkinson, D., Copeland, E.~J., 
\& Bassett, B.~A.\ 2004, \prd, 70, 083006 
\bibitem[Davis et al. 2007]{Davis2007} Davis, T. M. et al.\ 2007, \ ApJ., 666, 716D 
\bibitem[Douspis et al. 2003]{Douspis2003} Douspis, M., Riazuelo, A., Zolnierowski, Y., Blanchard, A., 2003, A\&A, 405, 409   
\bibitem[Douspis et 
al. 2008]{Douspis2008} Douspis, M., Zolnierowski, Y., Blanchard, A., \& Riazuelo, A.\ 2008, \aap, 488, 47 
\bibitem[Efstathiou 
\& Bond 1999]{EfstathiouBond} Efstathiou, G., \& Bond, J.~R.\ 1999, \mnras, 304, 75 
\bibitem[Eisenstein \& Hu 1998]{EisensteinHU}Eisenstein, D.J., Hu, W., 1998, ApJ, 496, 605
\bibitem[Eisenstein et al. (2005)]{E05} Eisenstein, D.~J., 
et al.\ 2005, \apj, 633, 560 
\bibitem[Ferramacho et 
al. 2009]{FBZ} Ferramacho, L.~D., Blanchard, A., \& Zolnierowski, Y.\ 2009, \aap, 499, 21 
\bibitem[Freedman et al. 2009]{Freedman2009} Freedman, W.~L., et 
al.\ 2009, \apj, 704, 1036 
\bibitem[Frieman et al. 2008]{FTH08} Frieman, J.~A., Turner, M.~S., \& Huterer, D.\ 2008, \araa, 46, 385 
\bibitem[Gannouji et al. 2006]{wcross1} Gannouji, R., 
Polarski, D., Ranquet, A., 
\& Starobinsky, A.~A.\ 2006, Journal of Cosmology and Astro-Particle Physics, 9, 16 
\bibitem[Hannestad  2005]{cs} Hannestad, S.\ 2005, \prd, 
71, 103519 
\bibitem[Ichikawa 
\& Takahashi 2006]{2006PhRvD..73h3526I} Ichikawa, K., \& Takahashi, T.\ 2006, \prd, 73, 083526 
\bibitem[Kowalski et al. 2008]{Kowalski2008} Kowalski, M., et al.\ 
2008, \apj, 686, 749 
\bibitem[Kessler et al. 2009]{Kessler2009} Kessler, R., et al.\ 
2009, \apjs, 185, 32 
\bibitem[Lampeitl et al. 2009]{Lampeitl2009} Lampeitl, H., et al.\ 
2009, arXiv:0910.2193 
\bibitem[Linder 2003]{CPL2}  Linder, E. \ 2003, PhRvL, 90i, 091301L
\bibitem[Linder
\& Huterer 2005]{2005PhRvD..72d3509L} Linder, E.~V., \& Huterer, D.\ 2005,
\prd, 72, 043509
\bibitem[Linden \& Virey 2008]{LindenVirey2008} Linden, S., \& Virey, J.-M.\ 2008, \prd, 78, 023526 
\bibitem[Maor et al. 2001]{2001PhRvL..86....6M} Maor, I., Brustein, R., 
\& Steinhardt, P.~J.\ 2001, Physical Review Letters, 86, 6
\bibitem[Nolta et al. 2009]{wmap5} Nolta, M.~R., et al.\ 2009, \apjs, 180, 296
\bibitem[Percival et al. 2007]{PercivalDR5} Percival, W.~J., et 
al.\ 2007, \apj, 657, 645 
\bibitem[Percival et al. 2009]{Percival2009} Percival, W.~J., et 
al.\ 2009, arXiv:0907.1660 
\bibitem[Perlmutter et al. 1999]{Perlmutter}Perlmutter, S., Aldering, G., Goldhaber, G. et al.\ 1999, ApJ, 517, 2, 565
\bibitem[Pierpaoli et al. 2001]{Pierpaoli} Pierpaoli, E., Scott, 
D., \& White, M.\ 2001, \mnras, 325, 77 
\bibitem[Ratra 
\& Peebles (1988)]{RP88} Ratra, B., \& Peebles, P.~J.~E.\ 1988, \prd, 37, 3406 
\bibitem[Reid et al. 2009]{Reid2009} Reid, B.~A., et al.\ 2009, 
arXiv:0907.1659 
\bibitem[Reichardt et al. 2008]{ACBAR} Reichardt, C.~L., et 
al.\ 2009, \apj, 694, 1200 
\bibitem[Riess et al. 2007]{Riess2007} Riess, A.~G., et al. 2007, ApJ, 659, 98R
\bibitem[Riess et al. 1998]{Riess98} Riess, A.~G., et al.\ 
1998, \aj, 116, 1009 
\bibitem[Shafieloo et al. 2009]{2009arXiv0903.5141S} Shafieloo, A., Sahni,
V., \& Starobinsky, A.~A.\ 2009, arXiv:0903.5141
\bibitem[S{\'a}nchez 
\& Cole 2008]{SanchezCole} S{\'a}nchez, A.~G., \& Cole, S.\ 2008, \mnras, 385, 830 
\bibitem[Tegmark et al. 2006]{T06} Tegmark, M., et al.\ 
2006, \prd, 74, 123507 
\bibitem[Virey et al. 2005]{Vireyetal} Virey, J.-M., Taxil, P., 
Tilquin, A., Ealet, A., Tao, C., \& Fouchez, D.\ 2005, \prd, 72, 061302 
\bibitem[Wood-Vasey et al. 2007]{W-V2007} W. M. Wood-Vasey et al. \ 2007, \ ApJ, 666, 694W \\


\end{thebibliography}

\end{document}